\documentclass[ amsmath,amssymb,aps,twocolumn]{revtex4-2}
\usepackage{graphicx}
\usepackage{dcolumn}
\usepackage{bm}

\usepackage[dvipdfmx,
colorlinks=true,
urlcolor=blue,
citecolor=blue,
linkcolor=blue,
hyperfootnotes=false]{hyperref}

\begin{document}

\title{Entanglement bipartitioning and tree tensor networks}

\author{Kouichi Okunishi${}^{1}$}
\author{Hiroshi Ueda${}^{2,3,4}$}
\author{Tomotoshi Nishino${}^5$}
\affiliation{${}^1$Department of Physics, Niigata University,  Niigata 950-2181, Japan.\email{okunishi@phys.sc.niigata-u.ac.jp}}
\affiliation{${}^2$Center for Quantum Information and Quantum Biology, Osaka University,  Toyonaka 560-0043, Japan. \email{ueda.hiroshi.qiqb@osaka-u.ac.jp}}
\affiliation{${}^3$JST, PRESTO, Kawaguchi 332-0012, Japan }
\affiliation{${}^4$Computational Materials Science Research Team, RIKEN Center for Computational Science (R-CCS), Kobe 650-0047, Japan.}
\affiliation{${}^5$Department of Physics, Graduate School of Science, Kobe University, Kobe 657-8501, Japan, \email{nishino@kobe-u.ac.jp}}
 
\date{\today}

\begin{abstract}
We propose the entanglement bipartitioning approach to design an optimal network structure of the tree-tensor-network (TTN) for quantum many-body systems.
Given an exact ground-state wavefunction, we perform sequential bipartitioning of spin-cluster nodes so as to minimize the mutual information or the maximum loss of the entanglement entropy associated with the branch to be bipartitioned.
We demonstrate that entanglement bipartitioning of up to 16 sites gives rise to nontrivial tree network structures for $S=1/2$ Heisenberg models in one and two dimensions.
The resulting TTNs enable us to obtain better variational energies, compared with standard TTNs such as uniform matrix product state and perfect-binary-tree tensor network.
\end{abstract}

\maketitle


\section{ Introduction \label{sec1}}

The tensor network is a universal theoretical framework for understanding entanglement structures in quantum many-body systems, statistical mechanics, quantum information as well as quantum gravities.\cite{JPSJ2022,Ran2020,Orus2019}
Also, it provides practical numerical tools for simulating various quantum/classical many-body systems.
In particular, tree tensor network (TTN) states\cite{TTN2006,Otsuka1996,Tagliacozzo2009,Murg2010,Nakatani2013} have been widely utilized in well-established tensor-network algorithms such as variational matrix product state (MPS) approaches\cite{White1992,Ostlund1995, PWFRG,Schollwock2011,iTEBD,VUMPS} and tensor renormalization groups\cite{TRG2007,HOTRG2012}.
The strong disorder renormalization group for random spin systems, which is a conventional real-space renormalization group based on the energy spectrum, can be regarded as a one-way algorithm based on the TTN framework\cite{SDRG1,Hikihara1999,Goldsborough2014,Seki2020}.
In the context of quantum computation, recently,  the MPS-type algorithm has attracted renewal interest by demonstrating its efficient simulation of noisy intermediate-scale quantum devices.\cite{Zhou2020}

The TTN algorithms mentioned above have been sophisticated through various applications and checks for practical problems so far.
However, the network structure of tensors, which is a primarily important factor in determining their efficiency and accuracy, has been basically designed by the  semi-empirical way based on the physical property of target systems.
How can we construct an optimal network structure of the TTN to describe a given quantum many-body state? 
Although several approaches to adjusting tree network structures have been tested\cite{Legeza2003,Murg2015,Legeza2015,Krumnow2016,Gunst2018,Henrik2019,Legeza2020,Li2022,Legeza2022,Hikihara2022},  a theoretical framework together with a practical procedure to extract an appropriate TTN structure from the exact wavefunction is necessary for thoroughly understanding the physics of tensor networks. 
Recently, the tensor-network representation of quantum states becomes massively relevant to a numerical simulation of quantum circuits.
A clear answer to the question about the TTN structure will be also essential in the context of quantum computation.

To be specific, let us discuss the binary TTN, which includes the MPS and the conventional real-space renormalization based on the perfect binary tree network. 
For an $N$ site system, in general, the number of possible binary tree networks is given by $\Omega_N=(2N-3)!!$, which increases much more rapidly with increasing $N$ than the exponential explosion of the Hilbert space dimension of the usual quantum many-body systems.\cite{Supp_1}
Then, our problem is to determine an appropriate TTN structure among such a huge network configuration space, which has not been systematically explored yet, even at the level of the binary TTN.
In this paper, we propose a top-down approach to determine the binary network structure of the TTN for a given ground-state wavefunction.

A central idea is that sequential bipartitioning of the exact ground-state wavefunction, which we call ``entanglement bipartitioning" (EBP) below,  generates an appropriate network structure so as to minimize the loss of entanglement entropy (EE) due to truncation of the bond dimension of isometry tensors. (See  Fig. 1)
In order to evaluate the loss of EE,  we examine the following complemental principles in the bipartitioning process: minimization of the mutual information between the branches bipartitioned and minimization of the maximum EE for the branches bipartitioned, which we respectively refer to as {\tt MMI} (minimization of mutual information) and as {\tt MMX} (minimization of the maximum loss) below.
We then show that both {\tt MMI} and {\tt MMX} generate the network structure corresponding to the dimer MPS for a $S=1/2$ Heisenberg chain of $16$ spins with the open boundary condition. 
Moreover, we find that the {\tt MMX} approach gives rise to a nontrivial network consisting of the 4-site MPS unit for a $S=1/2$ square-lattice Heisenberg model with the open boundary condition.
The variational optimization for the TTNs based on the EBP demonstrates that the resulting variational energies are certainly improved, compared with such well-known TTNs as uniform MPS and TTN with the perfect binary network structure (abbreviated as ``pbTTN'').
We also discuss relevance of the EBP to TTN simulations and its generalization to tensor networks including loop structures. 

This paper is organized as follows.
In the next section, we explain the details of the EBP with {\tt MMI} and {\tt MMX}.
In Sec. \ref{sec3}, we present numerical results of the EBP for $S=1/2$ Heisenberg spin chains and a $S=1/2$ square-lattice Heisenberg model with the open boundary condition.
We also evaluate variational energies for the resulting TTNs, using the direct variational optimization of isometry tensors with singular value decomposition (SVD) of environment tensors based on causal cone structure.\cite{Evenbly2009}.
In Sec. \ref{sec_4}, we summarize the results and discuss their implications to general tensor networks.

\section{ Entanglement bipartitioning \label{sec2}} 
 
\begin{figure}[tb]
\begin{center}
\includegraphics[width=8.2cm]{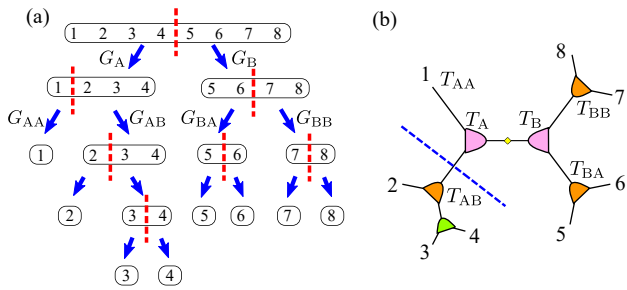}
\end{center}
\caption{(a) Schematic diagram of the EBP.
Given a wavefunction of $N=8$ (0th generation) , we search the bipartition of $G_\mathrm{A}$ and $G_\mathrm{B}$ minimizing $S(G_\mathrm{A})\, (=S(G_\mathrm{B}))$.
For the 1st generation, we further divide $G_\mathrm{A}\to G_\mathrm{AA}\,, G_\mathrm{AB}$ and $G_\mathrm{B}\to G_\mathrm{BA}\, , G_\mathrm{BB}$ by minimizing the evaluation function $f_\mathtt{MMI}$ or $f_\mathtt{MMX}$ based on the exact EE associated with subbranches.
(b) The TTN corresponding to (a), where $T_\mathrm{A}$, etc. represent isometry tensors. 
}
\label{fig1}
\end{figure}

As an example, we consider a $S=1/2$ Heisenberg chain of the length $N=8$ with open boundaries.
Suppose that the exact ground-state wavefunction $\Psi_{s_1, \cdots s_N}$ is obtained with such a numerical method as exact diagonalization, where $s_i$ with $i= 1, \cdots N$ indicates a spin state at $i$th site in the standard $S^z$ basis. 
As depicted in Fig. \ref{fig1}, we construct an appropriate network structure for the TTN from  entanglements involved in $\Psi_{s_1, \cdots s_N}$ (0th generation).
The first step is to divide $\Psi_{s_1, \cdots s_N}$ into two parts  $G_\mathrm{A}$ and $G_\mathrm{B}$ such that minimize $S(G_\mathrm{A})$, where $S(G_\mathrm{A}) (=S(G_\mathrm{B}))$ represents EE for the bipartitioned system.
In Fig. \ref{fig1}, we assumed $G_\mathrm{A} = \{1,2, 3,4\}$ and $G_\mathrm{B} = \{ 5, 6, 7, 8 \}$ (1st generation).

The next step is to divide the subsystems $G_\mathrm{A}$ and $G_\mathrm{B}$ into descendant subsystems of the 2nd generation: $G_\mathrm{A}\to G_\mathrm{AA},  G_\mathrm{AB}$ and $G_\mathrm{B}\to G_\mathrm{BA}, G_\mathrm{BB}$ respectively.
Suppose a sequence of the bipartitionings at the $n$th generation,  
\[\mathrm{Q}_{n} = \overbrace{\mathrm{AB\cdots A}}^{n}\]
for example. 
We then determine the optimal bipartitioning of $G_{\mathrm{Q}_{n}\mathrm{A}}, G_{\mathrm{Q}_{n}\mathrm{B}}$ for the $n+1$th generation.
For this purpose, we define the following function to evaluate the quality of bipartitioning based on the mutual information,
\begin{align}
f_{\tt MMI} & \equiv  S(G_{\mathrm{Q}_{n}\mathrm{A}})+ S(G_{\mathrm{Q}_{n}\mathrm{B}}) - S(G_{\mathrm{Q}_{n}\mathrm{A}}\cup G_{\mathrm{Q}_{n}\mathrm{B}}) .
\end{align}
We also examine a similar but distinct evaluation function,
\begin{align}
f_{\tt MMX}  & \equiv {\tt max}\big( S(G_{\mathrm{Q}_{n}\mathrm{A}}), S(G_{\mathrm{Q}_{n}\mathrm{B}})\big)  \, .
\end{align}
We then determine the optimal $G_{\mathrm{Q}_n\mathrm{A}}$ and $G_{\mathrm{Q}_n\mathrm{B}}$ so as to minimize $f_{\tt MMI}$ or $f_{\tt MMX}$.
In Fig. \ref{fig1}, we assume the case of $G_\mathrm{AA} = \{1\}$ and $G_\mathrm{AB} = \{ 2,3,4\}$, and $G_\mathrm{BA} = \{5,6\}$ and $G_\mathrm{BB} = \{7,8 \}$.
We can recursively repeat this process to the higher generations and finally arrive at the state where all of the spin sites in $\Psi_{s_1, \cdots s_N}$ are decomposed into single site nodes.
Then, the resulting binary tree network in Fig. \ref{fig1}(a) provides the corresponding TTN structure in Fig. \ref{fig1}(b);
More precisely, we put isometry tensors, which are illustrated as triangule-like symbols, on the nodes except for single-site and top nodes, and connect (contract) the tensor legs as in the network obtained by the EBP. 
Finally, we put a yellow diamond symbol at the top node representing the singular value matrix.

Let us discuss the physical implications of $f_{\tt MMI}$ and  $f_{\tt MMX}$ to the resulting TTN state.
First, we note that  $f_{\tt MMI}$ is nothing but the mutual information between two subsystems, implying that minimization of $f_{\tt MMI}$ basically leads to a classically well-separable bipartition, as partially used in Ref. \cite{Legeza2003}. 
For the TTN case, moreover,  $G_{\mathrm{Q}_{n}\mathrm{A}}\cup G_{\mathrm{Q}_{n}\mathrm{B}} = G_{\mathrm{Q}_n} $ and thus $S(G_{\mathrm{Q}_{n}\mathrm{A}}\cup G_{\mathrm{Q}_{n}\mathrm{B}}) ={\rm const}$ within the branch of $G_{\mathrm{Q}_{n}}$.
Thus, the minimum of $f_{\tt MMI}$ is determined by calculating $S(G_{\mathrm{Q}_{n}\mathrm{A}})$ and $ S(G_{\mathrm{Q}_{n}\mathrm{B}})$ only.

For $f_{\tt MMX}$, meanwhile,  we have more practical reasoning. 
In practical TTN simulations, we usually truncate leg degrees of freedom attached to isometric tensors, the upper bound of which is usually termed bond dimension $\chi$.
For a relatively small number of $\chi$, an approximated isometry with the bond dimension $\chi$ loses its accuracy by the truncation of the EE exceeding $\log \chi$. 
In particular, if the gap between the $\log \chi$ and the true EE without truncation for a certain isometry is large, the accuracy of its descendant branches is significantly spoiled. 
Thus, the minimization of $f_{\tt MMX}$ enables us to reduce the maximum loss of the EE due to truncation of the bond dimension in practical TTN simulations.
Here, we note that both approaches of ${\tt MMI}$ and ${\tt MMX}$ keep the balance of EEs between two branches at each bipartitioning step and thus basically generate the same binary tree network for one dimensional (1D) cases.
If an irregular difference occasionally appears in ramifications, however,  we adopt the {\tt MMX} approach, because it tends to pick up a more regular network structure.

In the EBP based on the above two approaches, the quantity we need is the exact EEs for all possible bipartitions in the subsystems.
The tree network structure generated by sequential bipartitioning might depend on branching paths.
However, we should note that the exact EE emerging at any link of tensors is independent of branching paths, if we keep sufficient bond dimensions.
More precisely, $S(G_{\mathrm{Q}_n \mathrm{A}})$ is exactly calculated as the bipartitioned EE of the $G_{\mathrm{Q}_n \mathrm{A}}$ and $\bar{G}_{\mathrm{Q}_n \mathrm{A}}$, where $\bar{G}_{\mathrm{Q}_n \mathrm{A}}$ is the complement of ${G}_{\mathrm{Q}_n \mathrm{A}}$. 
We also obtain the exact $S(G_{\mathrm{Q}_n \mathrm{B}})$ as the EE for the ${G}_{\mathrm{Q}_n \mathrm{B}}$ and $\bar{G}_{\mathrm{Q}_n \mathrm{A}}$, where $\bar{G}_{\mathrm{Q}_n \mathrm{B}}$ is the complement of ${G}_{\mathrm{Q}_n \mathrm{B}}$.
In Fig. \ref{fig1}(b), for example, the exact EE at the partitioning of the blue dotted line is calculated by the bipartition of $G_\mathrm{AB}=\{2,3,4\}$ and $\bar{G}_\mathrm{AB}=\{1,5,6,7,8\}$ for $\Psi_{s_1, \cdots s_8}$, which is clearly independent of the connectivity of the other links. 
Thus, we first calculate the EEs for all possible bipartitions, i.e. $2^{N-1}$ number of bipartitions of $\Psi_{s_1, \cdots s_N}$ beforehand, and then search the minimum of  $f_{{\tt MMI}/{\tt MMX}}$ by picking up the EEs for the corresponding bipartitions.
Here, we note that, in general, the bipartitions by minimizing $f_{{\tt MMI}/{\tt MMX}}$ may have trivial degeneracies reflecting symmetries such as lattice translation, parity, etc.
If this is the case, we randomly select one of the bipartitions having the degenerating value of $f_{{\tt MMI}/{\tt MMX}}$. Using the EBP, thus, we can determine an optimal network structure within a realistic computational cost up to $N=16$, without directly searching  $(2N-3)!!$ number of all possible networks.

Once the optimal network structure is generated by the EBP, we can straightforwardly construct the corresponding TTN by putting isometric 3-leg tensors at the nodes in the network (e.g. Fig. \ref{fig1}(b)). 
Here, it should be recalled that, in general, the location of the singular value tensor in the isometric TTN in the canonical form can be shifted with SVD or QR decomposition.\cite{TTN2006}
This implies that the location of the root node in Fig. \ref{fig1}(a) is irrelevant in the context of the TTN, and thus a binary tree network of the EBP accompanies a $2N-3$ number of equivalent TTN structures. 
In the following, we demonstrate variational optimization of the TTNs based on the EBP, where the resulting accuracy is confirmed to be independent of the position of top tensors.

\section{ Results and variational optimization \label{sec3}}

\begin{figure}[bt]
\begin{center}
\includegraphics[width=8cm]{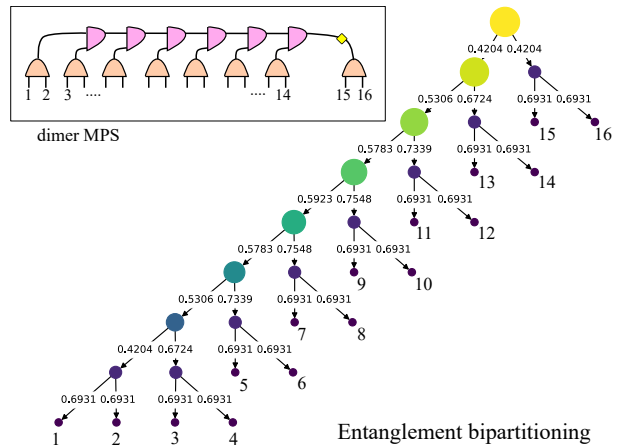}
\end{center}
\caption{The EBP for the ground state of a $S=1/2$ Heisenberg chain of $N=16$ sites with the open boundary condition. 
The size and color of node circles represent the number of spins included in the nodes.
Numerical values on link arrows represent the EE minimizing $f_{{\tt MMI}/{\tt MMX}}$.
Note that both $f_{{\tt MMI}/{\tt MMX}}$ yield the same network structure. 
Inset: The dimer MPS corresponding to the network structure in the main panel.}
\label{fig_2}
\end{figure}

\subsection{$S=1/2$ Heisenberg chain with the open boundary condition}
\label{ap_C}

As a benchmark example, we consider the $S=1/2$ Heisenberg chain of $N=16$ sites with the open boundary condition, for which the ground-state energy is given by $E_{\rm ex} = -6.911737146 \cdots $.
Here, we have assumed the exchange coupling, $J=1$. 
We perform the EBP based on $f_{{\tt MMI}/{\tt MMX}}$, which yields the same network structure depicted in Fig. \ref{fig_2}.
In this figure, we should note that the node circles represent not isometry tensors but nodes of the spin cluster, since the EBP generates the network structure only.
In particular, the size of circles in the figure symbolically represents the number of spins included in the nodes and numerical values on the links indicate optimal values of $S(G_{\mathrm{Q}_n\mathrm{A}}) $ and $S(G_{\mathrm{Q}_n\mathrm{B}})$ obtained with the EBP.  
Note that the decomposition of dimerized spin-pair nodes always gives the trivial EE, $\log 2\simeq 0.6931\cdots$, reflecting the spin-singlet ground state.

\begin{figure}[tb]
\begin{center}
\includegraphics[width=8cm]{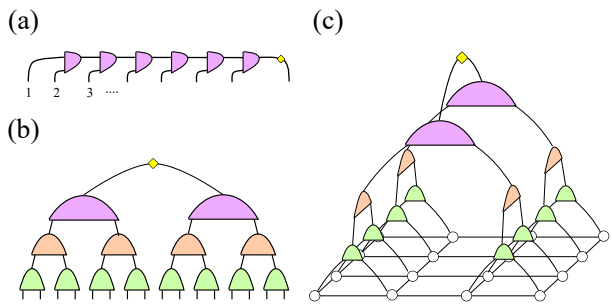}
\end{center}
\caption{(a) The uniform MPS, which is usually assumed in the DMRG.
(b) A TTN of the 1D perfect binary tree network, which is abbreviated as ``pbTTN".
(c) A pbTTN for a 2D square lattice.}
\label{fig_3}
\end{figure}

A crucial point in Fig. \ref{fig_2} is that spin pair nodes sequentially branch off from the dominant node clusters, which results in an MPS-type network structure with the dimerized-spin unit.
On the basis of the network structure in  Fig. \ref{fig_2}, we then construct the TTN state composed of 3-leg isometric tensors as in the inset of Fig. \ref{fig_2},  which is  mentioned as ``dimer MPS'' below.
In order to evaluate the quality of the network structure, we perform a variational calculation for the dimer MPS with the bond-dimension $\chi=8$. 
(See Appendix \ref{ap_B} for details of the variational optimization method.)
For comparison, we also examine variational optimizations of uniform MPS and pbTTN, which are respectively depicted in Fig. \ref{fig_3}(a) and (b).
Note that the ground-state energies are well converged within 200 iterations.
Results of the variational energy are summarized in Table \ref{tab_free}.
The dimer MPS gives the best energy  $E= -6.911614696$ in the table, which implies that the EBP approach generated an appropriate network structure for the spin chain with the open boundary condition.
Also, we have confirmed that the EEs hosted by the tensor legs in the dimer MPS are smaller than those in the uniform MPS.
Thus, the EBP certainly provide an appropriate network structure for the chain with open boundaries.

\begin{table}[bt]
\begin{center}
\begin{tabular*}{7.5cm}{r|ll}
\hline\hline
            & ~~~~~~~  $E$         & ~~~~~ $\Delta E$  \\ \hline
Exact       & ~~ -6.911737146$\cdots$ & ~~~~~ ---     \\ 
dimer MPS   & ~~ -6.911614696 & ~~0.00122     \\ 
uniform MPS & ~~ -6.911558558 & ~~0.00179     \\    
pbTTN       & ~~ -6.891960394 & ~~0.01977     \\ \hline\hline
\end{tabular*}
\end{center}
\caption{Variational energies for the $S=1/2$ Heisenberg chain of 16 sites with the open boundary condition.
The variational states are assumed to be dimer MPS [Inset of Fig. \ref{fig_2}], uniform MPS [Fig.\ref{fig_3}(a)] and pbTTN [Fig. \ref{fig_3}(b)].  The maximum bond dimension is $\chi=8$ for all cases.
$\Delta E$ denotes the deviation from the exact ground-state energy.}
\label{tab_free}
\end{table}

\subsection{ $S=1/2$ Heisenberg chain with the periodic boundary condition}
\label{ap_D}

\begin{figure}[bt]
\begin{center}
\includegraphics[width=7cm]{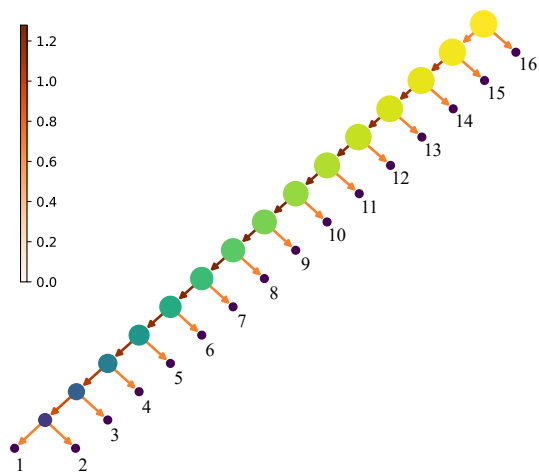}
\end{center}
\caption{The EBP for the $S=1/2$ Heisenberg chain of 16 sites with the periodic boundary, which yields the uniform chain network corresponding to the uniform MPS in Fig. \ref{fig_3}(a).
The color bar represents the magnitude of optimal values of the EE on link arrows. 
Note that the EE associated with the single-site bifurcation is always $\log 2$, reflecting the spin-singlet ground state.
}
\label{fig_4}
\end{figure}

We discuss the EBP for the ground state of the $S=1/2$ Heisenberg chain with the periodic boundary condition.
Of course,  the periodic boundary system is translationally invariant and thus the process of the EBP gives sequential single-spin decoupling with $S(G_{QB})=\log 2$, as depicted in Fig. \ref{fig_4}.
Then, the EBP naturally yields a uniform linear network, which leads us to a uniform MPS in Fig. \ref{fig_3}(a).
However, we also note that the resulting MPS explicitly breaks the translational symmetry at the boundary, implying that local tensors in  the vatiationally optimized MPS become position dependent. 
In strict sense, thus, the uniform MPS in Fig. \ref{fig_3}(a) may not mean ``uniform'' literally.

In Table \ref{tab_pbc}, we present variational energies of the chain with typical TTNs and the exact energy.
As in the case of the DMRG for the periodic-boundary system, the accuracy of the uniform MPS and dimer MPS for the periodic system is worse than that for the open boundary case, where we have confirmed that the bond energy between $i=1$ and $16$ lost its accuracy substantially.
Meanwhile, the pbTTN gives the best variational energy, which is comparable to the dimer MPS.
The reason why the pbTTN yielded a better accuracy is that it maintains more lattice symmetries such as bond parity involved in the ground-state wavefunction, which certainly mitigates a substantial drop in the accuracy of a particular bond energy compared with the uniform MPS.
This fact suggests that as for the variational TTN, the loop structure due to the periodic boundary condition is not treated appropriately in this EBP approach.
In the framework of the MPS, one should implement a periodic loop network structure to achieve an accurate variational energy for the periodic system.\cite{Verstraete2004}

\begin{table}[bt]
\begin{center}
\begin{tabular*}{7.5cm}{r|ll}
\hline\hline
            &   ~~~~~~~ $E$      & ~~~~~ $\Delta E$  \\ \hline
Exact       &   ~~  -7.142296361$\cdots$ & ~~~~~ ---     \\ 
dimer MPS   &   ~~  -7.106850777 & ~~ 0.03544558    \\ 
uniform MPS &   ~~  -7.095822585 & ~~ 0.04647378    \\    
pbTTN       &   ~~  -7.109020051 & ~~ 0.03327631    \\ \hline\hline
\end{tabular*}
\end{center}
\caption{Variational energies for the $S=1/2$ Heisenberg chain of 16 sites with the periodic boundary condition.
The variational states are assumed to be dimer MPS [Inset of Fig. \ref{fig_2}], uniform MPS [Fig.\ref{fig_3}(a)]and pbTTN [Fig. \ref{fig_3}(b)].  
The maximum bond dimension is fixed at $\chi=8$ for all cases.
$\Delta E$ denotes the deviation from the exact ground-state energy.}
\label{tab_pbc}
\end{table}

\subsection{$S=1/2$ square-lattice Heisenberg model with the open boundary condition}
\label{ap_E}

We discuss the EBP for 2D systems.
In general, the TTN does not satisfy the area law of EE for 2D systems.
This point should be contrasted to the tensor product state\cite{TPS} or projected entangled pair state\cite{PEPS}, which contain the loop structure with respect to the auxiliary degrees of freedom.
Here we restrict our argument to an optimal network structure within the binary tree network and then discuss features of the resulting TTN in comparison with typical TTNs, like a pbTTN and a MPS defined on snake-like 1D paths embedded in a 2D lattice.

\begin{figure}[tb]
\begin{center}
\includegraphics[width=8.2cm]{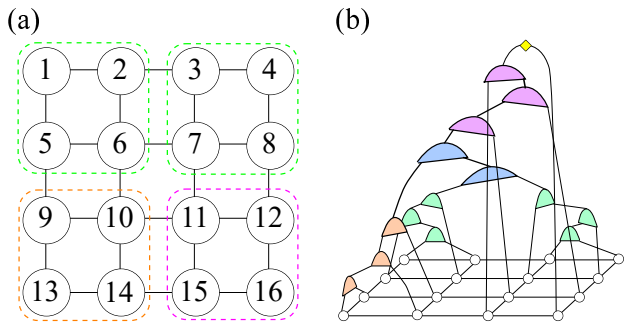}
\end{center}
\caption{The EBP for the Heisenberg model on a $4\times4$ square lattice. 
(a) Site indices on the square lattice.
Colored broken lines represent 4-site units generated by the EBP with $f_{\tt MMX}$.
(b) The TTN corresponding to the EBP result.} 
\label{fig_5}
\end{figure}

\begin{figure}[bt]
\begin{center}
\includegraphics[width=8cm]{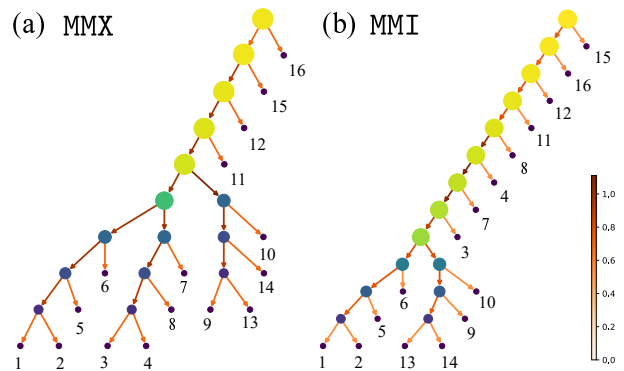}
\end{center}
\caption{The EBP for the square-lattice Heisenberg model of 16 sites with the open boundary condition.
The site index corresponds to that in Fig. 3(a).
The color bar represents the magnitude of optimal values of the EE on link arrows. 
Note that the EE associated with the single-site partitioning is always $\log 2$, reflecting the spin-singlet ground state.   
}
\label{fig_6}
\end{figure}

\begin{widetext}

\begin{figure}[bt]
\begin{center}
\includegraphics[width=15cm]{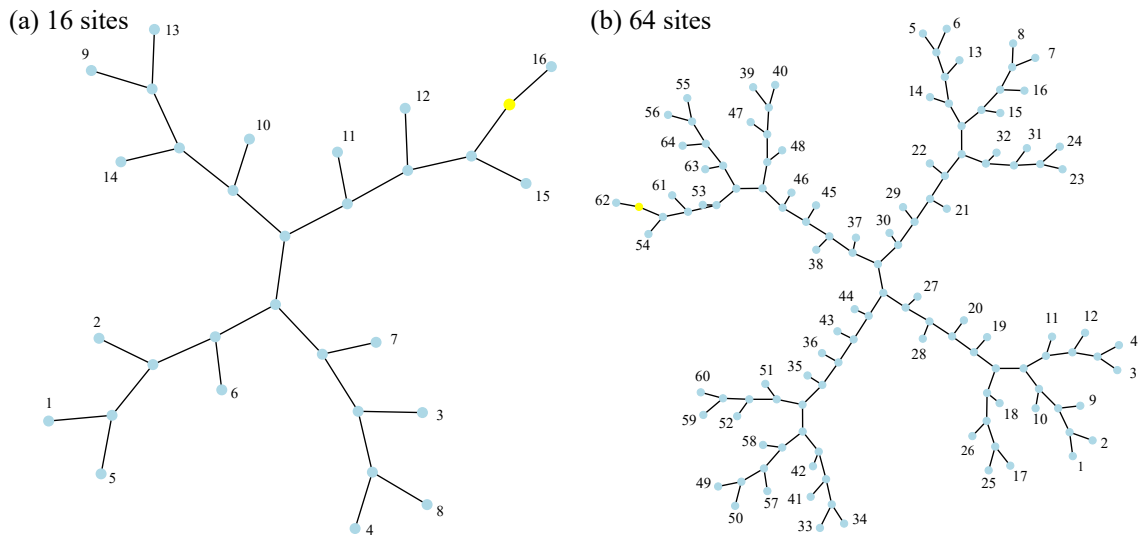}
\end{center}
\caption{The networks for the square-lattice Heisenberg model with open boundaries.
Numbers around the networks represent the index of bare-spin nodes.
(a) The network for the 16-site system in the disc-like representation,  which is equivalent to Fig. \ref{fig_6}(a) obtained with the EBP. The yellow node corresponds to the root(top) node in Fig. \ref{fig_6}(a) 
(b) The extended network for the 64-site system, which is obtained by a straightforward extrapolation of the panel (a).
}
\label{fig_7}
\end{figure}

\end{widetext}

We consider the EBP for the $S=1/2$ Heisenberg model on a square lattice of $4\times 4$ with the open boundary condition.
The arrangement of lattice sites is defined in Fig. \ref{fig_5}(a). 
We then perform the EBP for the exact ground-state wavefunction calculated with the exact diagonalization. 
The results of EBP with the {\tt MMX} and {\tt MMI} are presented in Fig. \ref{fig_6}, where the {\tt MMX} and {\tt MMI} approaches yield partly different network structures.  
In particular, the network structure due to $f_{\tt MMI}$ has some irregular bifurcations of distal branches, while the result of $f_{\tt MMX}$ shows a more systematic network structure consisting of the MPS-like network of the 4 spin unit, which is illustrated as broken lines in Fig. \ref{fig_5}(a) [See also Fig. \ref{fig_7}(a)].
Here, we note that the maximum EE in the bipartitioning process is $S_{\tt MMX}^\mathrm{(max)} = 1.111$ for the {\tt MMX}, while that for the {\tt MMI} is $S_{\tt MMI}^\mathrm{(max)} = 1.378$.
Thus, the loss of EE due to the truncation in the variational TTN computation for the {\tt MMX} network is expected to be smaller than that for the {\tt MMI}.
The TTN constructed on the {\tt MMX} network is shown in Fig. \ref{fig_5}(b), where the colors of isometries corresponds to those of the broken lines of the 4 spin unit in Fig. \ref{fig_5}(a).

In Table \ref{tab_square}, we present results of variational TTN calculations with $\chi=8$ based on the networks of the EBP, pbTTN, and snake MPS. 
The snake MPS has a typical 1D path of the MPS, which was often employed in the early works of the DMRG for 2D systems.
The pbTTN is depicted in Fig. \ref{fig_3}(c).
In the variational TTN optimization, the maximum bond dimension is fixed at $\chi=8$, and the maximum number of the TTN optimization is up to 200, where the numerical convergence of the variational energy is up to 7 digits.
In the table, the variational energy for the ${\tt MMX}$ is $E_{\tt MMX}=-9.052564$, which is close to the exact one $E_{\rm ex} = -9.189207\cdots$.
Also, the variational energy for the pbTTN for $N=16$ is the same as $E_{\tt MMX}$ within numerical convergence.
Thus, it may be concluded that the EBP with the {\tt MMX} yields the best variational TTN for a square-lattice Heisenberg model compared with such a typical TTN as snake-MPS.

\begin{table}[bht]
\begin{center}
\begin{tabular*}{7.5cm}{r|ll}
\hline\hline
           &   ~~~~~~~~ $E$  & ~~~~~~ $\Delta E$  \\ \hline
Exact      & ~~~ -9.189207$\cdots$   & ~~~~~~ ---     \\ 
{\tt MMX}  & ~~~ -9.052564   & ~~~ 0.136643    \\  
{\tt MMI}  & ~~~ -8.980623   & ~~~ 0.208584    \\  
pbTTN      & ~~~ -9.052564   & ~~~ 0.136643    \\ 
snake MPS  & ~~~ -8.760211   & ~~~ 0.428996    \\  
\hline\hline
\end{tabular*}
\end{center}
\caption{Variational ground-state energies for the square-lattice $S=1/2$ Heisenberg model of 16 sites with the open boundary condition.
The variational states of the {\tt MMX} and {\tt MMI} are respectively based on the networks in Fig. \ref{fig_6}(a) and (b). 
pbTTN indicates the variational energy for the TTN in Fig. \ref{fig_3}(c). 
Also, snake MPS corresponds to the MPS with a typical 1D path used in the DMRG for 2D systems.
The maximum bond dimension for variational computations is fixed to be $\chi=8$ for all cases.
$\Delta E$ denotes the deviation from the exact ground-state energy.}
\label{tab_square}
\end{table}

An interesting point in table \ref{tab_square} is that the variational energy of {\tt MMX} coincides with that of pbTTN.
Note that the coincidence of the variational energies for {\tt MMX} and pbTTN is confirmed for a larger $\chi$.
The main reason for this behavior is that the networks for both the TTNs have similar 4-site unit structures at the level of $N=16$.
As mentioned before, the position of root node (singular value tensor) is not relevant to the variational TTN computation. 
In the disk-like representation of the network in Fig. 7(a), which is equivalent to Fig. 5(c), we can move the yellow diamond symbol to the center bond in the network, which illustrate the 4-spin unit MPS structure similar to the pbTTN.  
The 4-spin unit MPS can describe the corresponding 4 spin cluster very accurately within the small $\chi$.
Thus, the resulting accuracy of the variational computation becomes equivalent to that for the pbTTN.
On the other hand, it remains a nontrivial problem how the variational energy for the TTNs behaves with increasing $N$.


In the remaining part of this subsection, we thus examine a straightforward extension of the {\tt MMX} network of $N=16(=4 \times 4)$ to $N=64(= 8 \times 8)$.
In particular, the result for $N=16$ also suggests that the {\tt MMX} network in Fig. \ref{fig_7}(a) may have better scalability with respect to $N$, whereas the EBP for $N>16$ is a numerically hard problem because of the computational limitation.
For this purpose, we notice that the network has a branching structure of a 4-site MPS-like unit, as is more visible in the disk-like representation of the network in Fig. \ref{fig_7}(a).
In Fig. \ref{fig_7}(b),  we then construct an extended network for the $N=64$($=8 \times 8$) site system by assuming self-similar connectivity in the network [See Fig. \ref{fig_7} (b)].

\begin{figure}[bt]
\begin{center}
\includegraphics[width=7cm]{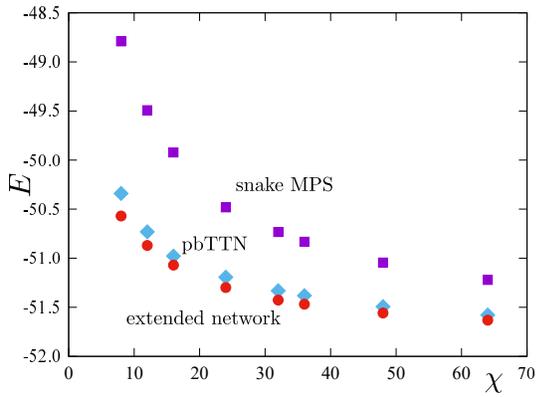}
\end{center}
\caption{The $\chi$-dependence of variational energies for the Heisenberg model on an $8\times 8$ lattice with the open boundary condition.
The extended network provides the best energy for each $\chi$. }
\label{fig_8}
\end{figure}

In order to evaluate the quality of the extended network, we perform variational TTN calculations for the corresponding TTN with $\chi=8 - 64$ as well as for the pbTTN and the snake-like MPS.
The $\chi$-dependence of variational energies is summarized in Fig. \ref{fig_8}, where the curves monotonously decrease with increasing $\chi$.
As expected from the result of $N=16$, then, it is verified that for each $\chi$, the {\tt MMX} always gives the lowest variational energy compared with those of the pbTTN and snake-like MPS.
The best value, $E=-51.63179$, is achieved by the extended-network with $\chi=64$, which is lower energy than $E_\mathrm{pbTTN} = -51.57896$ for the standard pbTTN and $E_\mathrm{snake}=-51.22124$ for the snake MPS often used in the DMRG.
We note that a computational time  was about 30 hours (37 hours) with Intel core-i9 12700K cpu for the extended network (pbTTN).
From Fig. \ref{fig_8}, moreover, it follows that the accuracy of the snake-like MPS with $\chi=64$ can be comparable to the TTN variation based on the extended network with $\chi\simeq 20$.
We therefore conclude that the EBP has certainly provided essential insight to construct a better network for the practical variational TTN simulation in two dimensions.

\section{Summary and discussions \label{sec_4}}

We have shed a light on the network structure of the TTN. 
We have proposed the EBP, which leads to an optimal binary network structure without using the full search of all possible binary TTNs.
After demonstrating the appropriate TTN for the $S=1/2$ Heisenberg chain with the open boundary, we have clarified that for the 2D Heisenberg model, a nontrivial hierarchical network with a MPS-like structure of the 4-spin unit provides better variational energy, compared with well-established TTNs such as snake MPS and pbTTN. 

From the practical viewpoint, designing an efficient network structure of the TTN is essential for reliable simulations of random spin systems\cite{SDRG1,Hikihara1999,Goldsborough2014,Lin2017,Seki2020,Seki2021,Ferrari2022}, quantum chemistry\cite{Murg2015,Krumnow2016,Gunst2018} and complex data processing\cite{Convy2022}, which usually contain highly-nonuniform interactions.
For such systems, the algorithm equipped with automatic structural optimization of the TTN is hopeful.\cite{Hikihara2022}  
We believe that the EBP provides an intriguing perspective of the variational TTN complemental to such a bottom-up construction of the TTN.

In this paper,  we have concentrated our arguments on the variational energy to evaluate the quality of the network structure.
However, the computational cost of variational optimization of TTNs is another important issue to be investigated from the computational viewpoint.
For example, let $d$ be the bond dimension of bare spins.
Then, the computational cost for a construction of the environment tensor based on the causal cone in Evenbly and Vidal's algorithm\cite{Evenbly2009} is ${\cal O}(d^2 \chi^4)$ for the MPS, which is much cheaper than  ${\cal O}(\chi^6)$  for such a generic TTN as pbTTN, if $d \ll \chi$.
This implies that the accuracy of an MPS simulation can be easily improved by increasing the bond dimension $\chi$.
In the practical situation, thus, the balance between the accuracy intrinsic to the network structure and the computational limitation to $\chi$ would become a crucial problem.

We finally point out that the {\tt MMI} approach has potential importance for analyzing general tensor networks containing loop structure. 
This is because $S(G_\mathrm{A}\cup G_\mathrm{B})$ in $f_{\tt MMI}$ is capable of representing interferences between the two branches mediated by a disentangler in the multi-scale entanglement renormalization ansatz (MERA)\cite{MERA2007} 
Recently, the nontrivial connection\cite{Swingle2012} between the MERA and the holographic entanglement entropy\cite{RT_PRL2006} has been stimulating intensive cross-disciplinary researches.
Thus, the EBP with {\tt MMI} may provide essential insight for developing a deeper understanding of the connection between tensor network structures and geometry associated with the holographic EE.\cite{cMERA2013,Nozaki2012,Caputa2017},
More recently, the TTN representation of quantum circuits is also discussed\cite{Seitz2022}.
The EBP approach may have relevance to designing quantum circuits.

\section*{Acknowledgment}

The authors thank T. Hikihara and K. Harada for valuable discussion.
The tensor contraction code used in this work was generated with TenosrTrace by G. Evenbly.\cite{tensortrace}
This work is partially supported by KAKENHI Grant Nos. JP21K03403, JP21H04446, JP22H01171 and  a Grant-in-Aid for Transformative Research Areas "The Natural Laws of Extreme Universe---A New Paradigm for Spacetime and Matter from Quantum Information" (KAKENHI Grant Nos. JP21H05182, JP21H05191) from JSPS of Japan.
It is also supported by MEXT Q-LEAP Grant No. JPMXS0120319794, JST PRESTO No. JPMJPR1911, and JST COI-NEXT No. JPMJPF2014. H.U and T.N was supported by the COE research grant in computational science from Hyogo Prefecture and Kobe City through Foundation for Computational Science.

\let\doi\relax

\bibliography{partition}

\newpage 
\appendix

\section{Number of  binary tree networks}
\label{ap_A}
The number of possible bipartitioned tree networks corresponds to that of rooted binary trees, which is given by
\begin{align}
\Omega_N = (2N -3)!!\, ,
\label{num_btn}
\end{align}
where $N(\ge 1)$ denotes the number of lattice sites (bare-spin nodes).
This formula can be derived from the recursive relation for $\Omega_N$,  
\begin{align}
\Omega_{N+1}= \sum_{k=1}^{\lfloor\frac{N+1}{2}\rfloor} {}_{N+1}C_n \frac{\Omega_k \Omega_{N-k+1}}{1+\delta_{N+1-k,k} }\, ,
\end{align}
which is associated with the trees of $k$ sites connecting at the rightmost branch of the trees of $N+1-k$ sites.
Note that Kronecker's $\delta$ in the denominator originates from the double counting of the same branches.
Then, this recursive relation can be attributed to the identity of double factrials\cite{double_fact},
\begin{align}
(2N -1)!! = \sum_{k=1}^{N} {}_NC_k \,(2k-3)!! (2N-2k-1)!!\, ,
\end{align}
which gives Eq. (\ref{num_btn}).
Note that $\Omega_{16} = 6190283353629375$.
Thus it is hard to directly search all possible binary networks even for $N=16$.

For the tree networks of $N$ sites, there are $2N-3$ number of possible edge positions for the root.
Thus, the number of the unrooted binary tree networks is given by
\begin{align}
\tilde{\Omega}_N = (2N -5)!!
\label{unrooted_btn}
\end{align}
with $\tilde{\Omega}_2 =1$.
In general, the position of the singular value tensor is not relevant in the TTN representation of quantum states.
Thus, one can regard  Eq. (\ref{unrooted_btn}) as the number of possible network structures for the TTN of $N$ sites.

\section{Variational optimization for the TTN}
\label{ap_B}

We explain details of variational optimization for the TTN and numerical results for the typical quantum spin systems. 
The algorithm used in this work is the direct variational optimization of isometry tensors through SVD of environment tensors based on causal cone structure.\cite{Evenbly2009}
The TTN contraction code for the environment tensors is generated by the TensorTrace package developed by G. Evenbly.\cite{tensortrace}
Initial tensors of an iterative optimization are random tensors, and the maximum bond dimension is fixed at $\chi=8$.
For the 1D chain of 16 spins, a typical number for iterations is 200, with which we can achieve 10-digits convergence of the variational energy.  
The variational optimization of the extended network for the 2D model of 64 spins with $\chi=64$ requires 400 iterations for 7 digits convergence.
Note that the reason why we used a relatively small $\chi$ is to make the cutoff effect on the resulting variational energy clear.
The variational optimization may be trapped at metastable states sometimes.
We perform the same computations with different initial random tensors 10 times and adopt the best energy as a final result for a target system.

\end{document}